\documentclass{cpbtex}
\usepackage{color}

\begin{document}
\begin{CJK*}{GBK}{song}

\title{Loss Mechanism Analyses of Perovskite Solar Cells with an Equivalent Circuit Model\thanks{\textcolor{red}{The paper is an English translated version of the original Chinese paper published in Acta Physica Sinica. Please cite the paper as: T. Xu, Z. S. Wang, X. H. Li, and W. E. I. Sha, Loss Mechanism Analyses of Perovskite Solar Cells with an Equivalent Circuit Model. Acta Physica Sinica 70: 098801 (2021). doi:10.7498/aps.70.20201975} }}

\author{Ting Xu$^{1}$, \ Zi-Shuai Wang$^{2}$, \ Xuan-Hua Li$^{3}$, \ and \ Wei E. I. Sha$^{1*}$\\
$^{1}${\small State Key Laboratory of Modern Optical Instrumentation,
}\\{\small College of Information Science \& Electronic Engineering, Zhejiang University, Hangzhou 310027, China}\\
$^{2}${\small Department of Electrical and Electronic Engineering, The University of Hong Kong, Hong Kong 999077, China}\\
$^{3}${\small State Key Laboratory of Solidification Processing, School of Materials Science and Engineering,}\\{\small Northwestern Polytechnical University, Xi'an 710072, China}\\
$^{*}${weisha@zju.edu.cn}} 


\date{}
\maketitle

\begin{abstract}
Perovskite solar cells have been attracting more and more attention due to their extraordinary performances in the photovoltaic field. In view of the highest certified power conversion efficiency of $25.5\%$ that is much lower than the corresponding Shockley-Queisser limit, understanding and quantifying the main loss factors affecting the power conversion efficiency of perovskite solar cells are urgently needed. At present, the three loss mechanisms generally recognized are optical loss, ohmic loss, and non-radiative recombination loss. Including the trap-assisted bulk recombination and surface recombination, the non-radiative recombination is proved to be the dominant recombination mechanism prohibiting the increase of efficiency. In this work, based on semiconductor physics, the expressions of bulk and surface recombination currents are analytically derived. Then taking the optical loss, series and shunt resistance losses, and bulk and surface recombination losses into consideration, an equivalent circuit model is proposed to describe the current density-voltage characteristics of practical perovskite solar cells. Furthermore, by comparing to the drift-diffusion model, the pre-defined physical parameters of the drift-diffusion model well agree with the fitting parameters retrieved by the equivalent circuit model, which verifies the reliability of the proposed model. For example, the carrier lifetimes in the drift-diffusion model are consistent with the recombination rates in the equivalent circuit model. Moreover, when the circuit model is applied to analyze experimental results, the fitting outcomes show favorable consistency to the physical investigations offered by the experiments. And the relative fitting errors of the above cases are all less than $2\%$. Through employing the model, the dominant recombination type is clearly identified and split current density-voltage curves characterizing different loss mechanisms are offered, which intuitively reveals the physical principles of efficiency loss. Additionally, through calculating the efficiency loss ratios under the open-circuit voltage condition, quantifying the above-mentioned loss mechanisms becomes simple and compelling. The prediction capability of the model is expected to be enhanced if a series of light intensity dependent current density-voltage curves are fitted simultaneously. Consequently, this model offers a guideline to approach the efficiency limit from a circuit-level perspective. And the model is a comprehensive simulation and analysis tool for understanding the device physics of perovskite solar cells.
\end{abstract}



\section{Introduction}
In recent years, the analysis of the loss mechanisms that greatly affect the power conversion efficiency of the perovskite solar cells (PVSCs) has attracted widespread attention from academia and industry, as the research on PVSCs progress. For an ideal single-junction perovskite solar cell, electrons and holes only recombine to emit photons through radiative recombination, and its theoretical efficiency limit, known as the Shockley-Queisser limit is about $31\%$$^{[1]}$. However, the highest power conversion efficiency that has been experimentally certified for practical PVSCs is $25.5\%$$^{[2]}$, which is still far from the theoretical limit.  On one hand, the radiative recombination of free carriers in perovskite is weak$^{[3]}$. On the other hand, various loss mechanisms tremendously affect the generation, transport and collection of carriers, and ultimately lead to the reduction in power conversion efficiency.

Based on the detailed balance theory, our previous work reported that for practical PVSCs, there are three main loss mechanisms that limit the power conversion efficiency, one is optical loss, the other is defect-assisted non-radiative (SRH) recombination loss, and the third is Ohmic loss$^{[4]}$. On the basis of the modified detailed balance model, taking the light trapping structure and the photon recycling effect into consideration, the major loss mechanisms affecting the operation of cells are quantified, and the relative fitting errors between theoretical and experimental current density-voltage curves ($J-V$ curves) are less than $4\%$.  However, the modified model neglects or cannot distinguish the effect of surface non-radiative recombination. For PVSCs, non-radiative recombination mechanisms have been proven to be the dominated recombination mechanism$^{[5,6]}$. At present, it is widely believed that the non-radiative recombination in PVSCs includes bulk SRH recombination, bulk Auger recombination and surface SRH recombination. Among them, due to low Auger recombination rate$^{[7]}$ of perovskite materials, the Auger recombination is often ignored when modeling the devices. Besides, the bulk recombination is mostly related to the inherent point defects and impurities, while the surface recombination is related to the surface defects in the perovskite layer$^{[8]}$. Therefore, judging the dominant non-radiative recombination mechanism of solar cells, and analyzing and quantifying the influence of bulk and surface recombination on the J-V curves are of great significance to the improvement of efficiency and stability for PVSCs$^{[9,10]}$.

For diagnosing non-radiative recombination types, the existing methods are generally divided into testing methods and simulation methods. The testing methods include ideality factor method$^{[11]}$ and perturbation method, such as impedance spectrum analysis$^{[12-14]}$, transient photovoltage measurement$^{[15,16]}$, and spectral measurement$^{[17]}$, etc. Unfortunately, the ideality factor method cannot analyze the impact of surface recombination. And the perturbation method is hard to quantify the bulk and surface recombination losses, and requires costly equipment support. As for the simulation approach, based on the drift-diffusion model, combined with the bulk and surface recombination formulae, the simulation model can be applied to emulate the $J-V$ curves. However, limited works on drift-diffusion model $^{[18]}$ discuss the optical loss and the connections between the detailed balance theory and the drift-diffusion model. Until 2017, applying the light-trapping and angular-restriction incorporated Roosbroeck-Shockley equation, and the selective electrode boundary conditions, Xingang Ren et al. obtained the equivalent conditions between the detailed balance theory and the drift-diffusion model. But with strong nonlinear characteristics, the drift-diffusion model is not suitable for fitting experimental $J-V$ curves, and thus is difficult to extract corresponding physical parameters. Given these, the drift-diffusion model is too complicated to understand the working mechanisms of the practical PVSCs.

In order to solve the above problems, based on the modified detailed balance model$^{[4]}$, and fully considering the optical loss, ohmic loss, bulk recombination loss and surface recombination loss, an improved equivalent circuit model is proposed. Based on the improved equivalent circuit model, the $J-V$ characteristics of PVSCs are described, and the efficiency loss mechanisms are analyzed.  Meanwhile, to verify the reliability of the model, simulated $J-V$ curves obtained by the drift-diffusion model and the experimental $J-V$ curves are compared to the fitted $J-V$ curves by the improved equivalent circuit model, respectively. Furthermore, through the extraction and numerical comparisons of characteristic parameters (series resistance, parallel resistance, bulk recombination factor and surface recombination factor), the ability of distinguishing non-radiative recombination types by the improved equivalent circuit model is tested, and the effect of quantifying each loss mechanism by the proposed model is evaluated.

\section{Theory}
\subsection{Equivalent circuit model}
Under illumination, an ideal photovoltaic cell can be regarded as a parallel circuit, which consists of an ideal diode and a constant current source, where the current $J_{ph}$ generated by the constant current source is called the photo-generated current. Based on the modified detailed balance model$^{[4]}$, the improved equivalent circuit model describing the $J-V$ curve of a practical PVSC is shown in Fig. 1.

\centerline{\includegraphics[width=120mm]{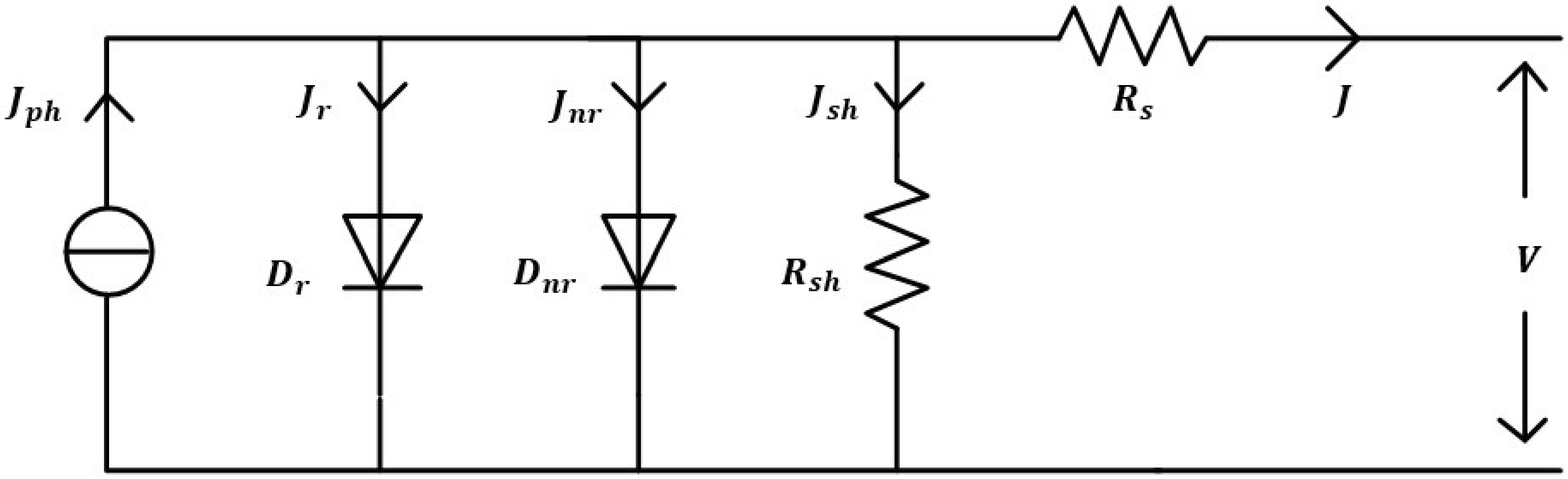}}
\vskip -6mm
\centerline{\footnotesize Fig.~1.~Equivalent circuit model of PVSC.}

\vskip 0.55\baselineskip

In the manufacturing process, practical solar cells will inevitably produce defects such as traps and pinholes, which always induce leakage current loss. This kind of loss is usually represented by the shunt resistance $R_{sh}$ in the circuit level. And the net effect of the ohmic loss at the anode and cathode electrodes, carrier transport layers and the interfaces of perovskite layer is commonly characterized by the series resistance $R_{s}$. And the ideal diode $D_r$ represents the radiative recombination of the perovskite, while the non-radiative recombination process is represented by the diode $D_{nr}$.

The $J-V$ characteristic of the equivalent circuit model is given by Eq. (1).
\begin{eqnarray}
J(V)=J_{ph}-J_r(V)-J_{bulk}(V)-J_{surf}(V)-J_{sh}(V),
\end{eqnarray}
where $V$ is the photovoltage of the solar cell, $J_r$ is the radiative recombination current caused by the photon recycling, $J_{bulk}$ is the non-radiative bulk recombination current, $J_{surf}$ is the surface recombination current, and $J_{sh}$ is the shunt resistance current (leakage current).

The photo-generated current is of the form:
\begin{eqnarray}
J_{ph}=q\int_0^{\infty}{\alpha \left( \lambda ,L \right) \frac{\varGamma \left( \lambda \right) \lambda}{hc_0}d\lambda},
\end{eqnarray}
where $c_0$ represents the speed of light, $\varGamma$ is the AM 1.5G solar spectrum, $\lambda$ represents the wavelength, and $q$ is the elementary charge. The absorptivity $\alpha$ is equal to the ratio of the power absorbed by the perovskite active layer and the incident power of the Sun, and depends on the thickness of the perovskite layer, the refractive index of the selected materials and the design of the light-trapping structure. Besides, the photocurrent can be solved by Maxwell's equations numerically.

Based on the detailed balance theory, radiative current is represented as
\begin{eqnarray}
J_r=q\int_0^{\infty}{\alpha \left( \lambda ,L \right) \frac{\varGamma _0\left( \lambda \right) \lambda}{hc_0}d\lambda}\left[ \exp \left( \frac{q\left( V+JR_s \right)}{k_BT} \right) -1 \right],
\end{eqnarray}
where $\varGamma _0$ is the blackbody radiation spectrum of the PVSC at $T=300\ K$, and $k_B$ is the Boltzmann constant.

Ignoring the Auger recombination, the dominant mechanisms of non-radiation recombination in PVSCs are bulk and surface recombination. Therefore, the bulk recombination current and the surface recombination current can be expressed separately as:
\begin{eqnarray}
J_{bulk}=qL\gamma _{bulk}n_i\exp \left( \frac{q\left( V+JR_s \right)}{2k_BT} \right),
\end{eqnarray}
\begin{eqnarray}
J_{surf}=qL_{surf}\gamma _{surf}\frac{n_i^2}{p_0^{h}}\exp \left( \frac{q\left( V+JR_s \right)}{k_BT} \right),
\end{eqnarray}
where $J_{bulk}$ represents the bulk recombination current, $J_{surf}$ represents the surface recombination current. $\gamma_{bulk}$ and $\gamma_{surf}$ are the bulk and surface recombination factors, respectively. $L$ is the thickness of the perovskite active layer, $L_{surf}$ is the effective thickness of the interfaces between the transport layers and the perovskite layer, $n_i$ is the intrinsic carrier density of the perovskite active material, and $p_0^h$ is the equilibrium majority (hole) carrier density on the perovskite side at the hole transport layer/perovskite layer interface (or the equilibrium majority (electron) carrier density on the perovskite side at the electron transport layer/perovskite layer interface $n_0^e$). The specific derivations of Eqs. (4, 5) can be found in Appendix A1.

The leakage current $J_{sh}$ is described by Eq. (6)
\begin{eqnarray}
J_{sh}=\frac{V+JR_s}{R_{sh}},
\end{eqnarray}

Based on the improved equivalent circuit model, by fitting the experimental $J-V$ curves of PVSCs, four parameters can be extracted, namely, the series resistance $R_s$, shunt resistance $R_{sh}$, bulk recombination factor $\gamma_{bulk}$ and surface recombination factor $\gamma_{surf}$. According to these parameters, the primary factors that lead to decreased efficiency can be analyzed, and the contributions of each loss mechanism (series resistance loss, shunt resistance loss, bulk recombination loss and surface recombination loss) to efficiency loss are discussed. In addition, by comparing the curves of $J_{bulk}-V$, $J_{surf}-V$ and $J_{sh}-V$, the evolution processes of bulk recombination, surface recombination and resistance loss can be effectively visualized with the change of voltage. Consequently, the model is able to reasonably conduct the analysis of the PVSCs' working mechanisms from the perspective of the circuit level, understand the loss mechanisms, and identify more precise and specific directions for further enhancement on the power conversion efficiency of PVSCs.

\subsection{Simulation methodology}

To check the reliability and accuracy of the equivalent circuit model for distinguishing the non-radiative recombination mechanisms, three $J-V$ curves obtained by the drift-diffusion model are adopted as the reference curves. Among them, the first $J-V$ curve represents the performance of the practical PVSC, in which bulk recombination is the dominant non-radiative recombination mechanism. The second $J-V$ curve describes the working characteristic of the PVSC when the surface recombination is the dominant non-radiative recombination type. And the third curve depicts the corresponding $J-V$ curve when there are no non-radiative recombination channels in the cell but the mobilities of the transport layers vary.

The drift-diffusion model describing the operating characteristics of the PVSC is governed by the Poisson equation, the drift-diffusion equations and the current continuity equations. The Poisson equation are shown in the following formula
\begin{eqnarray}
\frac{\partial}{\partial x}\left( \varepsilon \frac{\partial \psi}{\partial x} \right) =-q\left( p-n \right) ,
\end{eqnarray}

For electrons and holes, the drift-diffusion equations are expressed as
\begin{eqnarray}
J_n=-q\mu _nn\frac{\partial \psi}{\partial x}+qD_n\frac{\partial n}{\partial x}, \\
J_p=-q\mu _pp\frac{\partial \psi}{\partial x}-qD_p\frac{\partial p}{\partial x},
\end{eqnarray}

The current-continuity equations are
\begin{eqnarray}
\frac{\partial n}{\partial t}=\frac{1}{q}\frac{\partial J_n}{\partial x}+G-R, \\
\frac{\partial p}{\partial t}=-\frac{1}{q}\frac{\partial J_p}{\partial x}+G-R,
\end{eqnarray}
where $\psi$ is the potential, $q$ is the elementary charge, $n$ and $p$ are the electron density and hole density, respectively, $J_n$ and $J_p$ are the electron current density and hole current density, respectively. $\mu _n$, $\mu _p$, $D_n$, $D_p$ are the electron mobility, hole mobility, electron diffusion coefficient and hole diffusion coefficient, respectively. $G$ is the generation rate, and $R$ is the recombination rate.

The recombination mechanisms considered here consist of radiative recombination and non-radiative recombination. Non-radiative recombination involves bulk recombination and surface recombination. Therefore, the radiative recombination rate $R_{rad}$, bulk recombination rate $R_{bulk}$ and the surface recombination rate $R_{surf}$ are respectively as follows
\begin{eqnarray}
R_{rad}=k_{rad}\left( np-n_{i}^{2} \right), \\
R_{bulk}=\frac{np-n_{i}^{2}}{\tau _n\left( p+p_t \right) +\tau _p\left( n+n_t \right)}, \\
R_{surf}=\frac{n^+p^--n_{i}^{2}}{\tau _{surfn}\left( p^-+p_t \right) +\tau _{surfp}\left( n^++n_t \right)},
\end{eqnarray}
where $n_i$ is the intrinsic carrier density, $k_{rad}$ is the radiative recombination factor, $\tau _n$, $\tau _p$, $\tau _{surfn}$, $\tau _{surfp}$ are the bulk recombination lifetimes of electrons and holes, and the surface recombination lifetimes of electrons and holes, respectively. $n_t$, $p_t$ represent the concentration of trap electrons and holes. $n^+$, $p^-$ are the density of electrons and holes of the interfaces.

For reducing the influence of electrode surface recombination, the selective contact boundary conditions are
\begin{eqnarray}
J_{nc}=S_{nc}\left( n-n_{0c} \right), \quad S_{nc}=\infty , \\
J_{na}=S_{na}\left( n-n_{0a} \right), \quad S_{na}=0, \\
J_{pc}=S_{pc}\left( p-p_{0c} \right), \quad S_{pc}=0, \\
J_{pa}=S_{pa}\left( p-p_{0a} \right), \quad S_{pa}=\infty ,
\end{eqnarray}
where $S_{nc}$, $S_{na}$, $S_{pc}$, $S_{pa}$ are the surface recombination velocities of cathode electrons, anode electrons, cathode holes and anode holes, respectively. $n_{0c}$, $n_{0a}$, $p_{0c}$, $p_{0a}$ are the densities of cathode electrons, anode electrons, cathode holes and anode holes at the corresponding boundaries.

\section{Simulation analyses}

According to the drift-diffusion model in Section 2.2, three samples of cells are assumed. First we assume that there is only non-radiative bulk recombination in PVSC, that is, $\tau _{bulk}=100\ ns$, $\tau _{surf}=Inf$$^{[20]}$, majority carrier mobilities of transport layers are $\mu_{maj}=20\ cm^2/Vs$, minority carrier mobilities of transport layers are $\mu_{min}\approx0\ cm^2/Vs$$^{[21]}$. Then only surface recombination is assumed, in which let $\tau _{bulk}=Inf$, $\tau _{surf}=1\ ns$, $\mu_{maj}=20\ cm^2/Vs$, $\mu_{min}\approx0 \ cm^2/Vs$. And finally assume that there is no non-radiative recombination and the transport layers' mobilities are relatively low, that is, $\tau _{bulk}=Inf$, $\tau _{surf}=Inf$, $\mu_{maj}=1*10^{-2}\ cm^2/Vs$, $\mu_{min}=1*10^{-5}\ cm^2/Vs$. Without initial bias voltage, using the scanning rate of $0.1\ V/s$, corresponding simulated $J-V$ characteristic curves are modelled (scan from short circuit ($0\ V$) to slightly larger than open-circuit voltage ($V_{oc}$)), such as the red dotted lines in Fig. 2. Other simulation parameters used in the drift-diffusion model are presented as follows: device thickness $ETL\ 200\ nm/\ Absorber\ 500\ nm/\ HTL\ 200\ nm$$^{[22]}$, the relative dielectric constants of the electron transport layer, hole transport layer and active layer are respectively $\varepsilon _{r_{ETL}}=4$, $\varepsilon _{r_{HTL}}=4$, $\varepsilon _{r_{Absorber}}=31$$^{[23]}$, the mobilities of electrons and holes in the active layer are equal to $\mu _n=\mu _p=20\ cm^2/Vs$$^{[21]}$, the built-in electric field is $V_{bi}=0.8\ V$$^{[24]}$, and the band gap is $E_g=1.6\ eV$$^{[25,26]}$, the effective density of states of the perovskite layer are $N_c=N_v=10^{19}\ /cm^3$$^{[22]}$, the effective interface thickness between the perovskite layer and the transport layers is $L_{surf}=0.02\ nm$$^{[20]}$. Consequently, the intrinsic carrier density of the perovskite layer in the model is $n_i=\left( N_cN_v \right)^{1/2}\exp \left( -\frac{E_g}{2k_0T} \right) =4.336\times 10^5\ /cm^3$.

\vspace*{-4mm}

\centerline{\includegraphics[width=186mm]{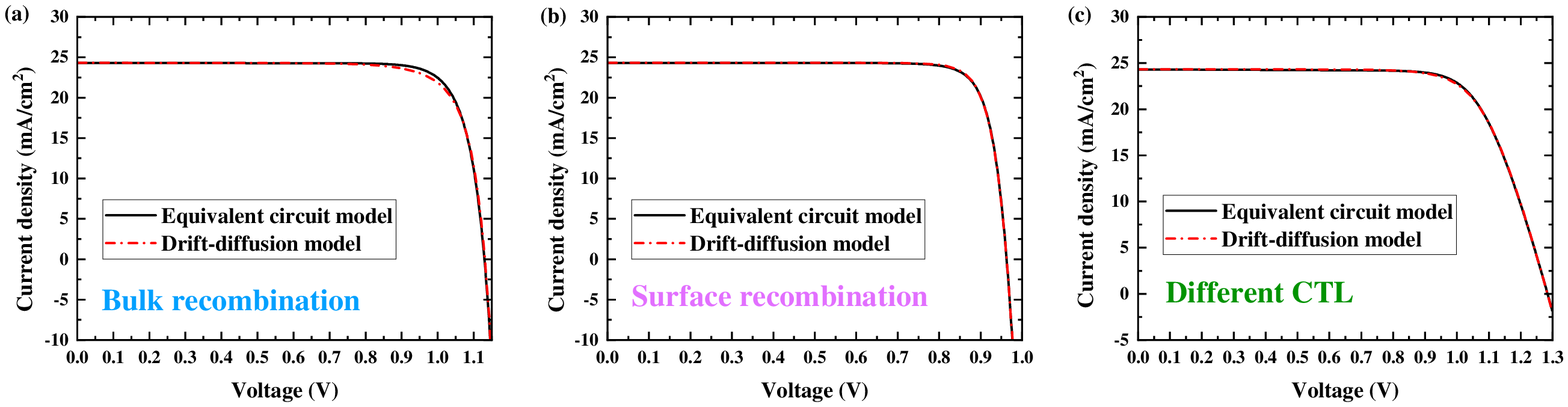}}
\vspace*{-4mm}

\centerline{\parbox[c]{16cm}{\footnotesize Fig.~2.~Current density-voltage curves of different non-radiative recombination types and different transport layers: (a) only bulk recombination is considered; (b) surface recombination is the dominant non-radiative recombination mechanism; (c) without non-radiative recombination and the mobilities of transport layers are changed. The red-dot lines represent $J-V$ curves that are simulated by drift-diffusion model, and the curves fitted by equivalent circuit model are shown in the dark solid lines.
}}

\vskip 0.55\baselineskip
\vspace*{2mm}

Based on the equivalent circuit model in Section 2.1, the reference curves containing only bulk recombination (red dot line in Fig. 2(a)), surface recombination (red dot line in Fig. 2(b)) and different transport layers (red dot line in Fig. 2(c)) are fitted (black curves), respectively. Obviously, good fitting results are obtained in the three cases. The relative fitting errors between the reference curves and the fitted curves are $1.01\%$, $0.2\%$, and $0.31\%$, respectively. It is clear that the extremely small fitting errors prove that the equivalent circuit model can describe the $J-V$ characteristics of PVSCs reliably and accurately.

At the same time, the fitting parameters are extracted and the power conversion efficiency analyses are performed in Table 1. Additionally, the drift-diffusion simulation shows that the equilibrium majority hole density close to the perovskite side at the hole transport layer/perovskite interface is $p_0^h=9.29*10^{10}\ /cm^3$, and the equilibrium majority electron density near the perovskite side at the electron transport layer/perovskite interface is $n_0^e=9.29*10^{10}\ /cm^3$.

\begin{center}
{\footnotesize{\bf Table 1.}The parameters retrieved from the $J-V$ curves of different cases.\\
\vskip 2mm \renewcommand{\arraystretch}{1.15} \tabcolsep 6pt

\begin{tabular}{ccccccccc}
\hline
{Cases} & {$\gamma_{bulk}$} & {$\gamma_{surf}$} & {$R_s$} & {$R_{sh}$} & {$J_{sc}$} & {$V_{oc}$} & {$FF$} & {$PCE$} \\
{ } & {$\left[s^{-1}\right]$} & {$\left[s^{-1}\right]$} & {$\left[Ohm\ cm^2\right]$} & {$\left[Ohm\ cm^2\right]$} & {$\left[mA\ cm^{-2}\right]$} &{$\left[V\right]$} &{$\left[\%\right]$ } &{$\left[\%\right]$} \\ \hline
{Bulk} &{$2.07*10^6$} & {$3.48*10^{5}$} & {$3.34*10^{-3}$} & {$1.46*10^{6}$} & {$24.28$} & {$1.13$} & {$82.33$} & {$22.58$} \\ \hline
{Surface} &{$1.30*10^7$} &{$1.95*10^{9}$} &{$3.84*10^{-1}$} &{$9.24*10^{6}$} &{$24.30$} &{$0.96$} &{$84.32$} &{$19.74$} \\ \hline
{CTL} &{$8.75*10^4$} &{$0.86$} &{$7.03*10^{-1}$} &{$7.00*10^{3}$} &{$24.32$} &{$1.28$} &{$73.15$} &{$22.85$} \\
\hline
\vspace{-0.5cm}
\end{tabular}}
\end{center}

\begin{center}
\parbox[c]{14.5cm}{\small Note 1: $\gamma_{srh}$ represents SRH bulk recombination factor, $\gamma_{surf}$ represents surface recombination factor, $R_s$ is series resistance, $R_{sh}$ is shunt resistance, $J_{sc}$, $V_{oc}$, $FF$ and $PCE$ represent the calculated short-circuit current, open-circuit voltage, fill factor and power conversion efficiency, respectively.\\}
\end{center}

Because of the equivalence between the drift-diffusion model and equivalent circuit model, for the same reference curves of PVSCs, the bulk non-radiative recombination factor and surface recombination factor are related to the bulk carrier lifetime and surface carrier lifetime, respectively. Furthermore, according to Appendix A1, $\tau=1/\gamma$, and thus calculate Table 2 to seek the relations between the non-radiative recombination factors and carrier lifetimes respectively from the two models.

\begin{center}
{\footnotesize{\bf Table 2.} The non-radiative recombination parameters retrieved from different cases by \\equivalent circuit model and drift-diffusion model.\\
\vskip 2mm \renewcommand{\arraystretch}{1.15} \tabcolsep 6pt
\begin{tabular}{ccccccc}
\hline
{Cases} & {$\tau_{bulk}$} & {$\frac{1}{\tau_{bulk}}$} & {$\gamma_{bulk}$} & {$\tau_{surf}$} & {$\frac{1}{\tau_{surf}}$} & {$\gamma_{surf}$} \\
{} &{$\left[s\right]$} &{$\left[s^{-1}\right]$} &{$\left[s^{-1}\right]$} &{$\left[s\right]$}
&{$\left[s^{-1}\right]$} &{$\left[s^{-1}\right]$} \\ \hline
{Bulk} &{$1.00*10^{-7}$} &{$1.00*10^{7}$} &{$2.07*10^{6}$} &{$Inf$} &{$Inf\ small$} &{$3.48*10^{5}$} \\\hline
{Surface} &{$Inf$} &{$Inf\ small$} &{$1.30*10^{7}$} &{$1.00*10^{-9}$} &{$1.00*10^9$ } &{$1.95*10^9$} \\\hline
{CTL} &{$Inf$} &{$Inf\ small$} &{$8.75*10^{4}$} &{$Inf$} &{$Inf\ small$} &{$0.86$} \\
\hline
\end{tabular}}
\end{center}

\vspace*{4mm}

For the PVSC device with only bulk recombination, $\tau_{bulk}=100\ ns$ and $\tau_{surf}= Inf$ are set in the drift-diffusion model. Hence the calculated value of the corresponding bulk recombination factor should be $10^7\ s^{-1}$, while the actual fitting value of the equivalent circuit model is $2.07*10^6\ s^{-1}$, which indicates that the fitting error of the bulk recombination factor in the case is within acceptable thresholds (about $1/5$ of the calculated value). And the calculated value of surface recombination factor $\gamma _{surf}$ should be infinitely small, but the fitting value of the practical equivalent circuit model is $3.48*10^5\ s^{-1}$. At this point, the fitting value of the surface recombination factor is significantly smaller than the fitting value of the bulk recombination factor (there's an order of magnitude difference), showing that the surface recombination in the PVSC is relatively weak. When only surface recombination involved, setting $\tau _{bulk}=Inf$, $\tau _{surf}=1\ ns$, the calculated value of the bulk recombination factor should be infinitely small, and the calculated value of the surface recombination factor should be $10^9\ s^{-1}$, while the fitting values of the equivalent circuit model are $\gamma _{bulk}= 1.30*10^7\ s^{-1}$, $\gamma _{surf}=1.95*10^{9}\ s^{-1}$. Similarly, in this case, the fitting error of the surface recombination factor is within the allowable range (about 2 times the calculated value), and the fitting value of the bulk recombination factor is significantly smaller than the fitting value of the surface recombination factor, demonstrating a weak bulk recombination feature. Moreover, for the device without non-radiative recombination and with lower transport layers' mobilities, the calculated bulk recombination factor and the surface recombination factor should both be infinitely small, however, the corresponding equivalent circuit fitting values are $\gamma_{bulk}=8.75*10^4\ s^{-1}$ and $\gamma_{surf}=0.86\ s^{-1}$, respectively. Compared to the previous two samples, it can be clearly seen that, in this case the fitting bulk recombination factor and the fitting surface recombination factor are relatively small, manifesting that the fitting non-radiative recombination factors and the setting lifetimes of the drift-diffusion model have good correspondences. In summary, given $L_{surf}$ and $p_0^{h}$, according to the comparisons of fitting $\gamma_{bulk}$ and $\gamma_{surf}$, we can roughly identify the non-radiative recombination situations in the PVSCs. But simply comparing the values of $\gamma_{bulk}$ and $\gamma_{surf}$ makes it inconvenient to intuitively understand the non-radiative recombination mechanisms in PVSCs. Consequently, a more vivid comparison method is needed.

As is shown in Table 1, the three cells have similar short-circuit currents. The maximum value of the open-circuit voltage is occurred when only surface recombination involved, while the maximum power conversion efficiency and lowest fill factor happened in the case without non-radiative recombination and with lower transport layers' mobilities. Besides, $R_s$ is always small and $R_{sh}$ is always large when only existing bulk recombination or surface recombination. Unlike the former features of resistances, when reducing the transport layers' mobilities, $R_s$ becomes larger and $R_{sh}$ becomes smaller. Similarly, simply observing the values of $R_s$ and $R_{sh}$, we can hardly ascertain the impacts of ohmic losses on the operating characteristics. Therefore, according to Eq. (1), the total current and sub-currents of PVSCs are calculated and drawn in Fig. 3.

For better understanding the roles of series resistance, shunt resistance, bulk recombination and surface recombination on the power conversion efficiency of PVSCs, a schematic diagram of the efficiency loss is given in Fig. 4 (see Appendix A2 for drawing methods).

Figures 3 and 4 unveil that regarding only bulk recombination or only surface recombination, the resistance currents in the device are rather small, and the efficiency losses caused by series and shunt resistance are both close to $0\%$. As is well-known, the main source of series resistance loss is from the ohmic loss of the transport layers and corresponding interfaces. The default mobilities in the drift-diffusion model corresponding to the two cases are $\mu_{maj}=20\ cm^2/Vs$, $\mu_{min}\approx0\ cm^2 /Vs$, indicating that both transport layers are high-conducting majority-carrier transport layers (minority-carrier blocking layers). Therefore, the series resistance loss of the cell should be extremely small, which is consistent with the calculated series resistance-induced efficiency loss ($0\%$). And when the mobilities of the transport layers are reduced, $\mu_{maj}=1*10^{-2}\ cm^2/Vs$ (the impedances of the transport layers increase), the series resistance loss increases to $42.16\%$. What's more, because the drift-diffusion model cannot capture the leakage current effect caused by the defects and pinholes (the leakage current is transmitted laterally and thus can not be collected by the upper and lower electrodes), the shunt resistance loss quantified by the equivalent circuit model should almost be $0\%$. In conclusion, the equivalent circuit model can well describe the contributions of resistances to $J-V$ characteristics.

\vspace*{1mm}
\centerline{\includegraphics[width=184mm]{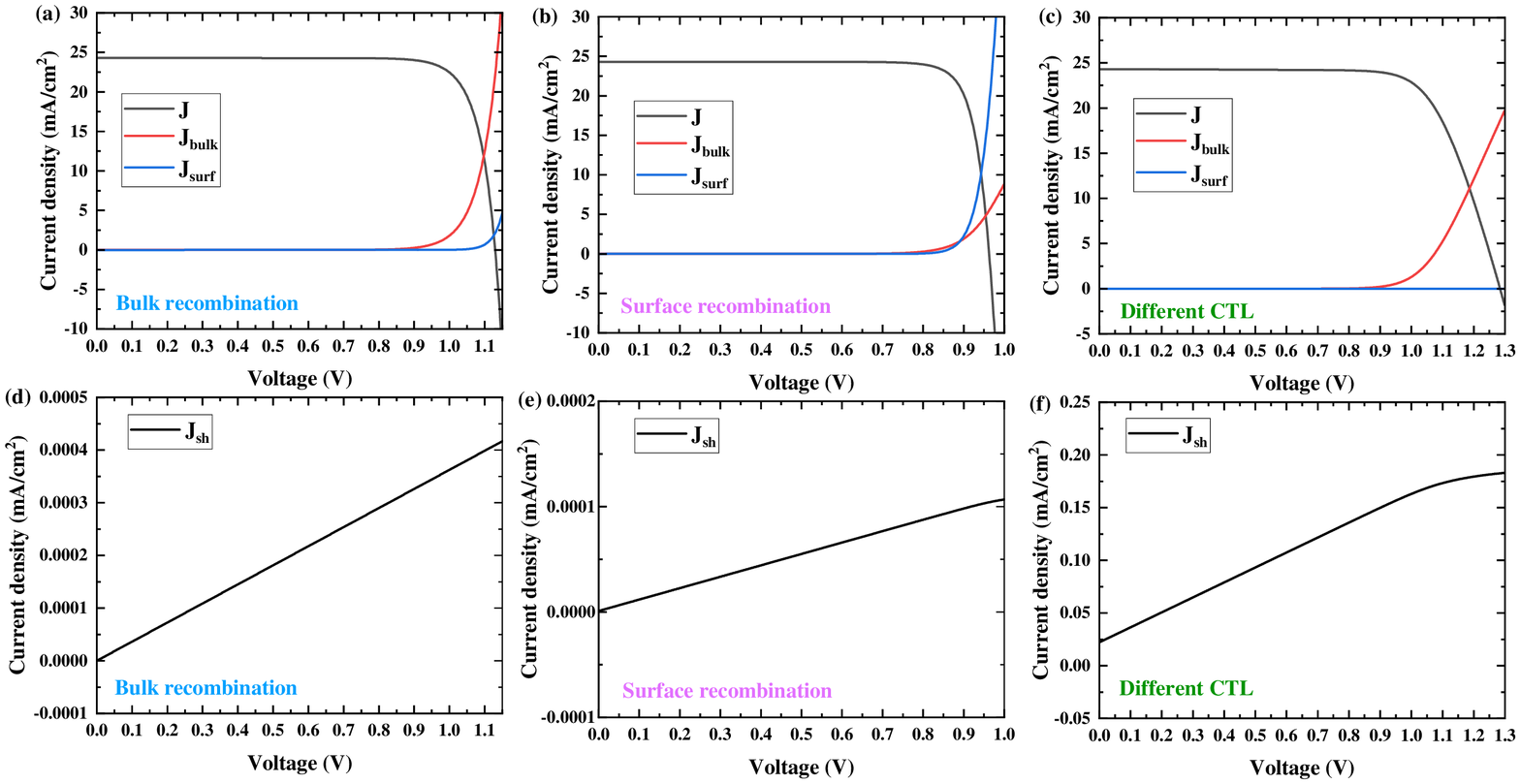}}
\centerline{\parbox[c]{16cm}{\footnotesize
Fig.~3.~Decompositions of the total current density of PVSCs according to Eq.(1): (a, d) only bulk recombination is considered; (b, e) only surface recombination is considered; (c, f) without non-radiative recombination and with different transport layers. (a, b, c) the total current, bulk recombination current and surface recombination current are described by black lines, red lines and blue lines, respectively; (d, e, f) $J_{sh}$ represents the (shunt) resistance current. }}

\vskip 0.55\baselineskip


\vspace*{0mm}

\vspace*{-5mm}
\centerline{\includegraphics[width=100mm]{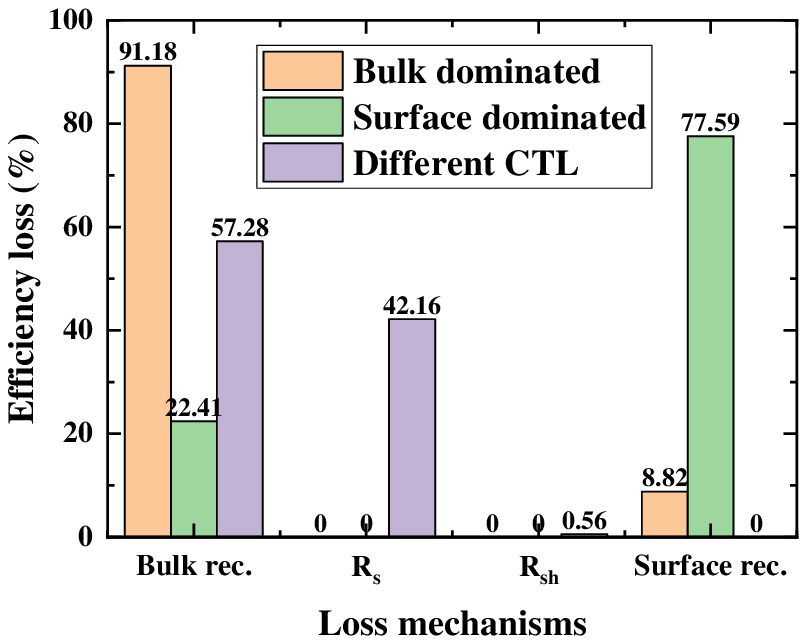}}
\vspace*{-7mm}
\centerline{\parbox[c]{9cm}{\footnotesize
Fig.~4.~The efficiency loss of perovskite solar cells in different cases. }}
\vskip 0.55\baselineskip

During the process of increasing the scanning voltage from $0\ V$ to $V_{bi}=0.8\ V$, the bulk current density $J_{bulk}$ and the surface current density $J_{surf}$ show little change (see Fig. 3(a)). As the voltage continues to scan from $V_{bi}$ to $V_{oc}$, both $J_{bulk}$ and $J_{surf}$ increase exponentially, but the starting voltage point of the exponential growth of $J_{surf}$ obviously lags behind $J_{bulk}$, resulting in more remarkable current growth of $J_{bulk}$ than $J_{surf}$. The dynamic behavior above-mentioned is basically in accordance with the preset conditions of the drift-diffusion model ($\tau_{bulk}=100\ ns$, $\tau_{surf}=Inf$). Figure 4 depicts that the calculated bulk recombination loss is $91.18\%$, which is much higher than the surface recombination loss of $8.82\%$, showing that the bulk recombination is dominant in the solar cell. Seen from Fig. 3(b), in the scanning process of $0\ V$ to $V_{bi}=0.8\ V$, $J_{bulk}$ and $J_{surf}$ also have no significant change. And during the process of $V_{bi}$ to $V_{oc}$, both $J_{bulk}$ and $J_{surf}$ increase exponentially from $V_{bi}$, but the growth rate of $J_{surf}$ is distinctly faster than that of $J_{bulk}$. Meanwhile, the bulk recombination loss accounts for $22.41\%$ of the total efficiency loss, and the surface recombination accounts for $77.59\%$ (see Fig. 4). Apparently, surface recombination becomes the major non-radiative recombination, and dominantly affects the state of efficiency loss when the PVSC operates. Again, this conclusion is consistent with the preset conditions of drift-diffusion model ($\tau _{bulk}=Inf$, $\tau _{surf}=1\ ns$).  For Fig. 3(c), $J_{bulk}$ still increases exponentially with the voltage starting from $V_{bi}$, but $J_{surf}$ keeps about $0\ mA/cm^2$. At the same time, the bulk recombination loss of $57.28\%$ is shown in Fig. 4, and the surface recombination loss is tiny. Corresponding to the drift-diffusion model, the reference cell has no non-radiative recombination and has different transport layers with lower mobilities, which indicates that there is almost no surface recombination in the cell. However, under extreme physical conditions, like no non-radiative recombination, the equivalent circuit model produces errors, which unfortunately induces extra bulk recombination behaviour in the simulation. Besides, the (series) resistance in this instance causes additional loss, which is in good agreement with the physical mechanism of the device. To sum up, the equivalent circuit model proposed can effectively distinguish the dominant non-radiative recombination mechanism of PVSCs, and quantify the impacts of different non-radiative recombination, series and shunt resistances, and then analyze the possible reasons for the disparities of the $J-V$ curves.

\section{Experimental results and analyses}

In this section, for further verifying the simulation ability of the proposed equivalent circuit model, and evaluating the validity of the quantitative ability of analyzing loss mechanisms, based on the reference [27], the measurement data of a PVSC under different grain boundary treatments are analyzed. According to the equivalent circuit model, the characteristic parameters extracted are shown in Table 3, where Control represents the MAPbI$_{3}$ PVSC device without the Lewis base or acid functional groups, DTS represents the $^{[28]}$ MAPbI$_{3}$ PVSC with DTS, while DR3T is MAPbI$_{3}$ PVSC with the BDT-based DR3TBDTT (abbreviated as DR3T)$^{[29]}$. In the simulation of the equivalent circuit model, it should be noted, since the $L_{surf}$ and $p_0^h$ are unknown in the PVSCs, $U_{surf}=\frac{L_{surf}\gamma _{surf}}{p_0^h}$ is introduced in Table 3 to describe the impact of surface recombination.

\begin{center}
{\footnotesize{\bf Table 3.} The parameters retrieved from the $J-V$ curves of different cases.\\
\vskip 2mm \renewcommand{\arraystretch}{1.15} \tabcolsep 6pt

\begin{tabular}{ccccccccc}
\hline
{Cases} &{$\gamma_{bulk}$} &{$U_{surf}$} &{$R_s$} &{$R_{sh}$} &{$J_{sc}$} &{$V_{oc}$} &{$FF$} & {$PCE$} \\
{} &{$\left[s^{-1}\right]$} &{$\left[nm\ cm^3/s\right]$} &{$\left[Ohm\ {cm}^2\right]$} &{$\left[Ohm\ {cm}^2\right]$}
&{$\left[mA\ cm^{-2}\right]$} &{$\left[V\right]$} &{$\left[\%\right]$} &{$\left[\%\right]$} \\ \hline
{ Control} &{$7.43*10^6$} &{$9.65*10^{-7}$} &{$2.10$} &{$1.73*10^{3}$} &{$21.29$} &{$1.06$} &{$76.03$} &{$17.24$} \\\hline
{ DTS} &{$1.89*10^6$} &{$8.61*10^{-7}$} &{$3.71$} &{$1.83*10^{3}$} &{$22.50$} &{$1.11$} &{$77.16$} &{$19.34$} \\\hline
{ DR3T} &{$7.17*10^5$} &{$1.96*10^{-6}$} &{$4.20$} &{$1.63*10^{3}$} &{$22.95$} &{$1.12$} &{$77.05$} &{$19.77$} \\
\hline
\end{tabular}}
\end{center}
\vspace{4mm}

According to Table 3, after the introduction of DTS, compared with Control, the bulk recombination factor of the PVSC is significantly reduced, and the open-circuit voltage is increased, indicating that DTS shows a good grain boundary passivation effect for MAPbI$_{3}$ active layer, and can effectively improve the working performance of the PVSC. Similarly, after introducing DR3T, the bulk recombination factor is further reduced, and the open-circuit voltage is further increased, manifesting that the grain boundary passivation effect of DR3T is outperform than DTS; but due to the increase of surface recombination, the efficiency enhancement is inapparent.

\vspace*{0mm}
\centerline{\includegraphics[width=184mm]{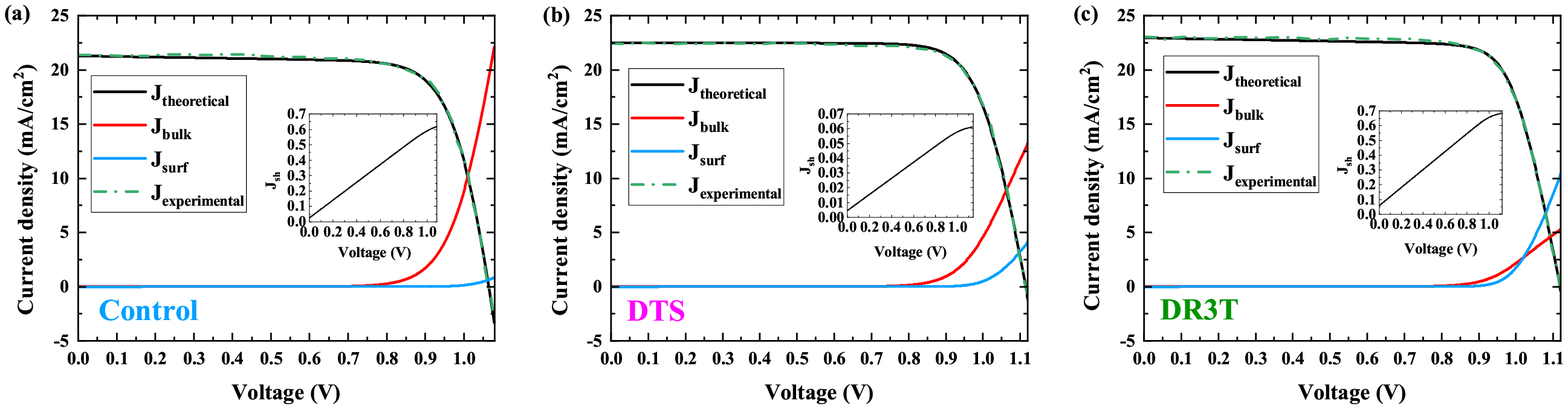}}
\vspace*{-4mm}
\centerline{\parbox[c]{17cm}{\footnotesize
Fig.~5.~Decompositions of the total current density of PVSCs according to Eq.(1): (a) devices based on the control MAPbI$_3$ films; (b) devices based on the DTS passivated MAPbI$_3$ films; (c) devices based on the DR3T passivated MAPbI$_3$ films. (a, b, c) the total theoretical current, bulk recombination current, surface recombination current and experimental current are described by solid black lines, red lines, blue lines and dotted cyan lines, respectively. The insets show the bias voltage dependence of $J_{sh}$}}

\vskip 0.55\baselineskip
\vspace*{3mm}

The fitted $J-V$ curves of the equivalent circuit model under different grain boundary treatments are depicted in Fig. 5. The pictures are explained as follows: Figure 5(a) represents the $J-V$ curves with no grain boundary modification, involving the experimental and fitted curve by the equivalent circuit model and the sub-currents of the total current. The relative fitting error of the total theoretical curve and the experimental curve is $1.08\%$.  Figure 5(b) shows the curves when DTS is introduced, and the relative fitting error of its theoretical and experimental curves is $0.70\%$.  Figure 5(c) depicts the curves when DR3T is introduced, and the relative fitting error is $0.95\%$. The extremely small relative fitting errors once again prove the excellent descriptions of the equivalent circuit model, when reproducing the $J-V$ curves of the practical PVSCs. And the corresponding schematic diagram of efficiency loss is presented in Fig. 6.

\vspace*{2mm}

\vspace*{0mm}
\centerline{\includegraphics[width=80mm]{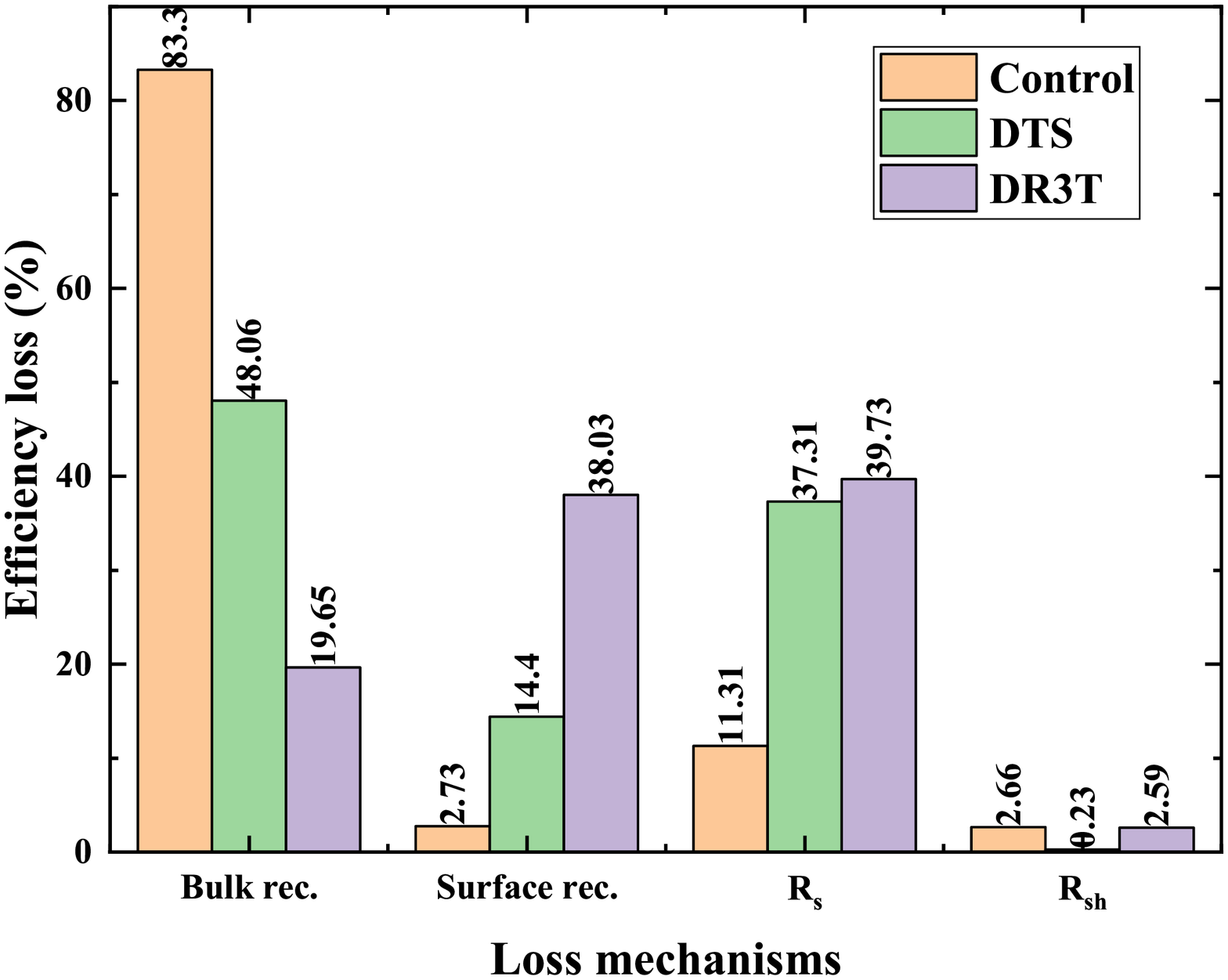}}
\vspace*{0mm}
\vskip -2mm
\centerline{\parbox[c]{9cm}{\footnotesize
Fig.~6.~The efficiency loss of PVSCs with different grain boundaries}}
\vskip 0.55\baselineskip
\vskip 4mm

According to Figs. 5 and 6, in PVSCs without grain boundary engineering, bulk recombination is the dominant non-radiative recombination mechanism, which causes an efficiency loss of $83.3\%$. For device with DTS, the bulk recombination loss is reduced to $48.06\%$, and the surface recombination loss increases to $14.40\%$ (when judging the non-radiative recombination characteristics of PVSCs under different conditions, the analysis of non-radiative recombination factors is not accurate enough. Hence the corresponding efficiency loss ratios are required). After introducing DR3T, the surface recombination is further increased to $38.03\%$, and the bulk recombination is further reduced to $19.65\%$, meaning that the surface recombination turns to the dominant non-radiative recombination mechanism of the PVSC. Refering to the analyses in literature [27], the interaction between DTS or DR3T molecules and perovskite is the cause of defect passivation and inter-grain carrier transport. Therefore, DTS and DR3T passivated cells possess the reduced bulk recombination loss. And because MAPbI$_3$ has a larger energetic disparity with DTS, DTS will thus hinder the transport of carriers to a certain extent, that is, increase the series resistance loss. What's more, DR3T should make the defect states in the perovskite layer shallower, which is helpful to the collection of electrons and holes, but also more liable to suffer larger surface recombination loss. In conclusion, the simulation results of the equivalent circuit model are basically consistent with the analyses of the literature. Therefore, the equivalent circuit model proposed can accurately determine the dominant non-radiative recombination type in practical PVSCs. And through the comparisons of the sub-currents and the calculations of the efficiency loss ratios, the model offers a better understanding of the working mechanisms and could design optimization strategies of PVSCs.

\section{Conclusion}

An improved equivalent circuit model is introduced to describe the current density-voltage characteristics of PVSCs. Involving photon recycling, light-trapping structure, non-radiative recombination (bulk recombination and surface recombination), series and shunt resistance losses, the proposed model is a rather comprehensive simulation tool for PVSCs modeling. Furthermore, to test and verify the accuracy of describing the $J-V$ curves by the proposed circuit model, reference curves simulated by the drift-diffusion model in conjugation with bulk and surface recombination formulae, and the experimental $J-V$ curves under different grain boundary passivation treatments are compared to the equivalent circuit model method. The relative fitting errors are within $2\%$. Based on the improved equivalent circuit model, by fitting the $J-V$ curves, the bulk recombination factor representing the effect of bulk recombination mechanism, the surface recombination factor indicating the influence of surface recombination, and series and shunt resistances of ohmic losses can be extracted. Given that, the loss factors that affect efficiency can be quantified. And additionally, through drawing bulk recombination current-density curves, surface recombination curves and resistance curves, the various losses during the voltage scanning process can be analyzed separately. Our work helps to identify the dominant loss mechanism and clarify corresponding working principle of PVSCs, so as to accurately identify the key point of efficiency optimization approach.

To approach the Shockley-Queisser theoretical limit, through simulation and analyses of the $J-V$ curve with only bulk recombination, curve dominated by surface recombination, and curve without non-radiative recombination and with changed transport layers, our theoretical results show that suitable blocking layers can significantly reduce the series resistance loss. Besides, optimal optical design, high quality of the perovskite active layer and passivated interface defects can effectively reduce non-radiative recombination loss and shunt resistance loss, hence obtain higher power conversion efficiency and fill factor$^{[30]}$. What's more, the model needs to further improve the uniqueness of the fitting parameters. Simultaneous fitting of $J-V$ curves under different light intensities can be applied to solve the uniqueness problem. Neglecting the abnormal hysteresis effect, the model is either incapable of explaining the impact of ion migration under different scanning conditions.

\section*{Acknowledgment}

The project is funded by the General Program of National Natural Science Foundation of China (Grant No. 61975177).

\newpage
\section*{Appendix A1}

When $n\approx p$, $n\gg n_t$ and $p\gg p_t$, the equation of bulk SRH recombination rate is given$^{[32]}$:
\begin{eqnarray}
R_{bulk}=\frac{np-n_{i}^{2}}{\tau _n\left( p+p_t \right) +\tau _p\left( n+n_t \right)}\approx \frac{n}{\tau _n+\tau _p},
\end{eqnarray}
where $n$ is the electron density, $p$ is the hole density, $n_i$ is the intrinsic carrier density of perovskite active layer, $p_t$ is the trap hole concentration, $n_t$ is the trap electron concentration, $\tau_n$ is the bulk recombination lifetime of electrons, and $\tau_p$ is the bulk recombination lifetime of holes.

The product of nonequilibrium carrier densities is
\begin{eqnarray}
np\approx n^2=n_{i}^{2}\exp \left( \frac{E_{Fn}-E_{Fp}}{k_BT} \right)
\end{eqnarray}
among them, $E_{Fn}$ is the quasi-Fermi level of electrons, and $E_{Fp}$ is the quasi-Fermi level of holes. If the carrier mobility is large enough, $\frac{E_{Fn}-E_{Fp}}{q}$ is the voltage applied at both ends of the perovskite layer $V_{ap}$.

Therefore, the bulk recombination current is
\begin{eqnarray}
J_{bulk}=qR_{bulk}L=qL\gamma _{bulk}n_i\exp \left( \frac{qV_{ap}}{2k_BT} \right)
\end{eqnarray}
where $q$ is the elementary charge, $L$ is the thickness of the perovskite layer, $\gamma_{bulk}$ is the bulk recombination factor, $k_B$ is Boltzmann constant, and $T$ is the temperature.

Under the conditions of $p^-\gg n^+$, $n^+\gg n_t$ and $p^-\gg p_t$, for the hole transport layer/perovskite layer interface, the surface SRH recombination rate is known as:
\begin{eqnarray}
R_{surf}=\frac{n^+p^--n_{i}^{2}}{\tau _{surfn}\left( p^-+p_t \right) +\tau _{surfp}\left( n^++n_t \right)}\approx \frac{n^+}{\tau_{surfn}},
\end{eqnarray}
where $n^+$ is the non-equilibrium minority electron density at the interface near the perovskite side, $p^-$ is the non-equilibrium majority hole density at the interface near the transport layer side, $\tau_{surfn}$ is the surface recombination lifetime of electrons, $\tau_{surfp}$ is the surface recombination lifetime of holes.

According to Boltzmann statistics, the minority carrier density $n^+$ is
\begin{eqnarray}
n^+=n_0^{h}\exp \left( \frac{qV_{ap}}{k_BT} \right)=\frac{n_{i}^{2}}{p_0^{h}}\exp \left( \frac{qV_{ap}}{k_BT} \right)
\end{eqnarray}
where $n_0^{h}$ and $p_0^{h}$ are the equilibrium minority electron density and the equilibrium majority hole density at the hole transport layer/perovskite layer interface, near the perovskite side, respectively. Besides, changing the doping density of the hole transport layer, and the barrier height relative to the valence band of the active layer will both affect $p_0^{h}$.

Similarly, at the interface of the electron transport layer and perovskite layer, the non-equilibrium minority hole density near the perovskite side is
\begin{eqnarray}
p^-=p_0^{e}\exp \left( \frac{qV_{ap}}{k_BT} \right)=\frac{n_{i}^{2}}{n_0^{e}}\exp \left( \frac{qV_{ap}}{k_BT} \right)
\end{eqnarray}
where $n_0^{e}$ and $p_0^{e}$ are the equilibrium majority electron density and the equilibrium minority hole density near the perovskite side at the electron transport layer/perovskite layer interface, respectively.

Consequently, the surface recombination current at the hole transport layer/perovskite interface, and the surface recombination current at the electron transport layer/perovskite interface are presented as follows:
\begin{eqnarray}
J_{surfh}=qL_{surfh}\gamma _{surfn}n^+=qL_{surfh}\gamma _{surfn}\frac{n_{i}^{_2}}{p_0^{h}}\exp \left( \frac{qV_{ap}}{k_BT} \right)
\end{eqnarray}
\begin{eqnarray}
J_{surfe}=qL_{surfe}\gamma _{surfp}p^-=qL_{surfe}\gamma _{surfp}\frac{n_{i}^{_2}}{n_0^{e}}\exp \left( \frac{qV_{ap}}{k_BT} \right)
\end{eqnarray}
where $L_{surfh}$ is the thickness of the hole transport layer/perovskite layer interface, and $L_{surfe}$ is the thickness of the electron transport layer/perovskite layer interface.

Assuming that $\gamma _{surfn}\approx \gamma _{surfp}$, $n_0^{e}\approx p_0^{h}$, and the effective thickness of the transport layers/perovskite layer is $L_{surf}$, then
\begin{eqnarray}
J_{surf}=qL_{surf}\gamma _{surf}\frac{n_{i}^{_2}}{p_0^h}\exp \left( \frac{qV_{ap}}{k_BT} \right)
\end{eqnarray}

\section*{Appendix A2}

For the ideality factor extraction method, to avoid the impact of parasitic resistances, Kristofer et al. select the open-circuit voltage $V_{oc}$ as the function of light intensity to improve the accuracy of the ideality factor extraction when judging the dominant non-radiative recombination mechanism$^{[11]}$. Similarly, without current inside the PVSC at the open-circuit voltage point, the impacts of non-radiative bulk recombination, non-radiative surface recombination, series resistance and shunt resistance on the point can be analyzed separately and more reliably. Then compare above four sets of power to the ideal power, the efficiency loss ratios of PVSCs caused by the four mechanisms can therefore be quantified.

Take DR3T in Section 4 as an example. By fitting the equivalent circuit model, the open-circuit voltage is known ($V_{oc}=1.12\ V$).

Firstly, let the bulk recombination factor $\gamma_{bulk}=0\ s^{-1}$, surface recombination factor $\gamma_{surf}=0\ s^{-1}$, series resistance $R_s=0\ Ohm\ cm^2$ and shunt resistance $R_{sh}=Inf\ Ohm\ cm^2$, and draw the $J-V$ curve as the black line in Fig. 7. And meanwhile, calculate the ideal power of the PVSK marked as $P_{id}=18.18\ mW/cm^2$ when $V=1.12\ V$.

Secondly, let the bulk recombination factor $\gamma_{bulk}=7.17*10^5\ s^{-1}$, surface recombination factor $\gamma_{surf}=0\ s^{-1}$, series resistance $R_s=0\ Ohm\ cm^2$ and shunt resistance $R_{sh}=Inf\ Ohm\ cm^2$, and draw the $J-V$ curve as the red line in Fig. 7. And at the same time, calculate the corresponding power of the PVSK marked as $P_{bulk}=12.36\ mW/cm^2$ when $V=1.12\ V$.

Thirdly, let $\gamma_{bulk}=0\ s^{-1}$, $U_{surf}=1.96*10^{-6}\ s^{-1}$, $R_s=0\ Ohm\ cm^2$, $R_{sh}=Inf\ Ohm\ cm^2$, and draw the $J-V$ curve as the blue line in Fig. 7. And calculate the corresponding power of the PVSK marked as $P_{surf}=6.93\ mW/cm^2$ when $V=1.12\ V$.

Fourthly, let $\gamma_{bulk}=0\ s^{-1}$, $U_{surf}=0\ s^{-1}$, $R_s=4.20\ Ohm\ cm^2$, $R_{sh}=Inf\ Ohm\ cm^2$, and draw the $J-V$ curve as the pink line in Fig. 7. And at the same time, calculate the corresponding power of the PVSK marked as $P_{Rs}=6.42\ mW/cm^2$ when $V=1.12\ V$.

Finally, let $\gamma_{bulk}=0\ s^{-1}$, $U_{surf}=0\ s^{-1}$, $R_s=0\ Ohm\ cm^2$, $R_{sh}=1.63*10^{3}\ Ohm\ cm^2$, and draw the $J-V$ curve as the cyan line in Fig. 7. And meanwhile, calculate the corresponding power of the PVSK marked as $P_{Rsh}=17.41\ mW/cm^2$ when $V=1.12\ V$.

Consequently, on the basis of the relations between the above power and the ideal power, the effects of bulk recombination factor, surface recombination factor, series resistance and shunt resistance on the efficiency loss can be quantified, respectively.

\vspace*{0mm}

\vspace*{0mm}
\centerline{\includegraphics[width=90mm]{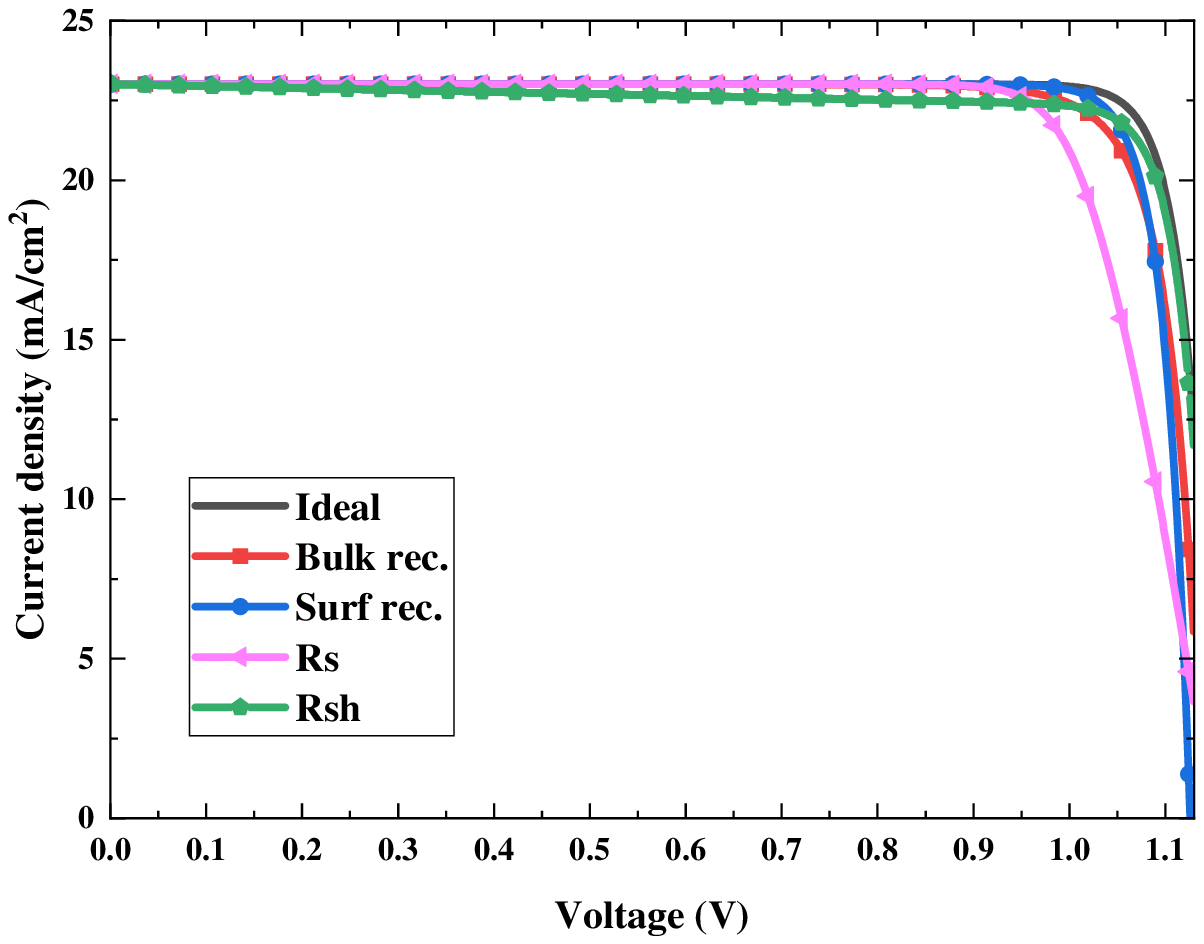}}
\vspace*{0mm}
\vskip -6mm
\centerline{\parbox[c]{7cm}{\footnotesize
Fig.~7.~The method of quantifying efficiency loss of PVSCs.}}
\vskip 0.55\baselineskip
\vskip 6mm


\end{CJK*} 
\end{document}